\documentclass[prl,aps,showpacs,onecolumn]{revtex4}
\usepackage{graphicx}
\usepackage{bm}

\begin{document}

\title{Semiclassical Dynamics of Dirac particles interacting with a Static
Gravitational Field}
\author{Pierre Gosselin$^{1}$, Alain B\'{e}rard$^{2}$ and Herv\'{e} Mohrbach$^{2}$}

\address{$^1$ Universit\'e Grenoble I, Institut Fourier, UMR
5582 CNRS-UJF, UFR de Math\'ematiques, BP74, 38402 Saint Martin
d'H\`eres, Cedex, France \\
$^2$ Universit\'e Paul Verlaine, Institut de Physique, ICPMB1-FR
CNRS 2843, Laboratoire de Physique Mol\'eculaire et des
Collisions, 1, boulevard Arago, 57078 Metz, France}

\begin{abstract}
The semiclassical limit for Dirac particles interacting with a static
gravitational field is investigated. A Foldy-Wouthuysen transformation which
diagonalizes at the semiclassical order the Dirac equation for an arbitrary
static spacetime metric is realized. In this representation the Hamiltonian
provides for a coupling between spin and gravity through the torsion of the
gravitational field. In the specific case of a symmetric gravitational field
we retrieve the Hamiltonian previously found by other authors. But our
formalism provides for another effect, namely, the spin hall effect, which
was not predicted before in this context.
\end{abstract}

\maketitle


\section{Introduction}

The dynamics of quantum particles in a gravitational field is an important
topic, at the crossroad of particle Physics and cosmology. Among the various
approaches to this problem (see for instance \cite{SOURIAU}), the use of the
Dirac equation in an external gravitational field (\cite{OBUKHOV}, \cite
{SILENKO}) has proven to be useful to obtain the semi classical corrections
to the electron dynamics. However the results obtained in the litterature
through a genuine Foldy-Wouthuysen transformation were restricted to a
certain class of weak static gravitational field \cite{SILENKO} and did not
give any information about the dynamics in a general gravitational field.

Moreover, these previous works did not take into account the role of Berry
phases, which has recently proven to be very relevant in the context of the
semi classical dynamics. Actually, since the seminal work of Berry \cite
{BERRY}, it is well known that a wave function acquires a geometric phase
when a quantum mechanical system has an adiabatic evolution. But it is only
recently that a possible influence of the Berry phase on the transport
properties (in particular on the semiclassical dynamics) of several physical
systems has been investigated. For instance, in \cite{ALAIN} and \cite
{BLIOKH1} the adiabatic evolution of the Dirac electron in an external
potential was investigated. In \cite{PIERRE}, the role of the Berry Phase on
the Dynamics of an electron in a periodic potential has been explicitly
derived in an Hamiltonian approach, in accord with the independent works of
\cite{NIU} and \cite{DUVAL}. In our previous works, it was shown that
position and momentum operators acquire spin-orbit contributions which turn
out to be a Berry connections rendering the algebraic structure of the
coordinates and momenta non-commutative (as found also in \cite{DUVAL}).
This drastically modifies the semiclassical equations of motion and implies
a topological spin transport similar to the intrinsic spin Hall effect in
semiconductor \cite{ZHANG}.

All these results presented similar patterns in the resulting coordinate
algebra and equations of motion and it was shown in \cite{Methode} that they
all fall in a general formalism of semiclassical diagonalization for a large
class of quantum system, including the Dirac Hamiltonian. This method has
the advantage to derive directly the role of the Berry phase from the
diagonalization procedure and introduces naturally the role of
noncommutative dynamical variables.

In this paper, we adapt the formalism developed in \cite{Methode} to the
case of a Dirac electron in an arbitrary static gravitational field. We
find, at order $\hbar $, the Foldy-Wouthuysen diagonalization for the
Hamiltonian and assess the relevance of the Berry phases in the procedure.
Our approach generalizes the results of \cite{OBUKHOV} and \cite{SILENKO}
(these last approaches, although different, being physically equivalent)
since our method has also the advantage to be valid for all kinds of static
gravitational fields and does not require a small field expansion. Moreover,
we show that the Hamiltonian includes corrections similar to spin orbit and
orbital momentum coupling found in \cite{BLIOKH1} for Dirac electrons in
electromagnetic fields. Here, the role of the magnetic field is played by
the torsion of the gravitational field.

We then derive the ''covariant'' coordinates algebra and their non
commutative commutation relations. These variables differ from the ordinary
position and momentum by the Berry terms and lead to different equation of
motion than the usual ones \cite{SILENKO}. We ultimately derive the semi
classical equations of motion for the electron, showing the appearance at
order $\hbar $ of new contributions to the force equation.

\section{Electron in a Static Gravitational Field}

Consider an electron propagating in a static gravitation field where the
corresponding metric satisfies $g_{0\alpha }=0$, which implies $%
ds^{2}=g_{00}(dx^{0})^{2}-g_{ij}dx^{i}dx^{j}$. In the sequel, we will denote
$\sqrt{g_{00}}=V(\mathbf{R})$, to be consistent with the notation of \ (\cite
{SILENKO}, \cite{OBUKHOV}). Following \cite{Leclerc} the Dirac Hamiltonian
for an electron in such a gravitational field has the form :
\begin{equation}
\hat{H}=V(\mathbf{R})\mathbf{\alpha }.\mathbf{\tilde{P}}+\frac{\hbar }{4}%
\varepsilon _{\varrho \beta \gamma }\Gamma _{0}^{\varrho \beta }(\mathbf{R})%
\mathbf{\Sigma }^{\gamma }+i\frac{\hbar }{4}\Gamma _{0}^{0\beta }(\mathbf{R}%
)\alpha ^{\beta }\text{\ }+V(\mathbf{R})\beta m
\end{equation}
and $\mathbf{\tilde{P}}$ given by $\tilde{P}_{\alpha }\mathbf{=}h_{\alpha
}^{i}(\mathbf{R})(P_{i}+\hbar \varepsilon _{\varrho \beta \gamma }\frac{%
\Gamma _{i}^{\varrho \beta }}{4}(\mathbf{R})\sigma ^{\gamma })$ with $%
h_{\alpha }^{i}$ the static orthonormal dreibein $(\alpha =1,2,3)$, $\Gamma
_{i}^{\alpha \beta }$ the spin connection components and $\varepsilon
_{\alpha \beta \gamma }\sigma ^{\gamma }=\frac{i}{8}(\gamma ^{\alpha }\gamma
^{\beta }-\gamma ^{\beta }\gamma ^{\alpha }).$ It is known \cite{Leclerc}
that for a static gravitational field (which is the case considered here),
the Hamiltonian $\hat{H}$ is hermitian.

Because the components of $\mathbf{\tilde{P}}$ depend both on the operators $%
\mathbf{P}$ and $\mathbf{R}$ the diagonalization at order $\hbar $ is
performed by adapting the method detailed in \cite{Methode} to block
diagonal Hamiltonians. To do so, we first write $\hat{H}$ in a symmetrical
way in $\mathbf{P}$ and $\mathbf{R}$. This is easily achieved using the
Hermiticity of the Hamiltonian: \ $\hat{H}=\frac{1}{2}\left( V(\mathbf{R})%
\mathbf{\alpha }.\mathbf{\tilde{P}+\tilde{P}}^{+}\mathbf{.\alpha }V(\mathbf{R%
})\right) +\frac{\hbar }{4}\varepsilon _{\varrho \beta \gamma }\Gamma
_{0}^{\varrho \beta }(\mathbf{R})\mathbf{\Sigma }^{\gamma }+V(\mathbf{R}%
)\beta m$. The diagonalization is then performed in two steps, similarly to
the method exposed in \cite{Photon}.

\subsection{Diagonalization when $\mathbf{P}$ and $\mathbf{R}$ commute}

We consider a formal situation where $\mathbf{R}$ is first considered as a
parameter $\mathbf{r}$ commuting with $\mathbf{P}$. Some computations show
that the Hamiltonian $\hat{H}_{0}$ (we add the index $0$ when $\mathbf{R}$
is a parameter) can then be diagonalized, at first order in $\hbar $ by the
following unitary FW matrix
\[
U_{0}(\mathbf{\tilde{P})}=D\left( E_{0}+V(\mathbf{r})m+c\frac{1}{2}\beta
\left( V(\mathbf{r})\mathbf{\alpha }.\mathbf{\tilde{P}+\tilde{P}}^{+}\mathbf{%
.\alpha }V(\mathbf{r})\right) +N\right) \mathbf{/}\sqrt{2E_{0}\left(
E_{0}+mV(\mathbf{r})\right) }
\]
with $E_{0}=\sqrt{\left( \frac{V(\mathbf{r})\mathbf{\alpha }.\mathbf{\tilde{P%
}+\tilde{P}}^{+}\mathbf{.\alpha }V(\mathbf{r})}{2}\right) ^{2}+m^{2}V^{2}(%
\mathbf{r})}$, $N=\frac{\hbar }{4}\frac{i\mathbf{\alpha }.\left( \mathbf{%
P\times \Gamma }_{0}\right) }{E_{0}}$, $D=1+\ \frac{\hbar }{4}\beta \frac{%
\left( \mathbf{P\times \Gamma }\right) \times \mathbf{P}}{2E_{0}^{2}(E_{0}+m)%
}$ and $\Gamma _{0\gamma }=\varepsilon _{\varrho \beta \gamma }\Gamma
_{0}^{\varrho \beta }(\mathbf{r})$.

The proof of this diagonalization relies on the fact that for each parameter
$\mathbf{r}$ the matrices $h_{\alpha }^{i}$ and $\Gamma _{i}^{\alpha \beta }$
are independent of both the momentum and position operators, $\beta $ and $%
\mathbf{\alpha .\tilde{P}}$ anticommute and in the Taylor expansion of $%
E_{0} $ all terms commute with $\beta $ and $\mathbf{\alpha .\tilde{P}+%
\mathbf{\tilde{P}}^{+}\mathbf{.\alpha }}$. In this context the diagonalized
Hamiltonian is equal to $U_{0}\hat{H}_{0}U{}^{+}=\beta \sqrt{\left( \frac{V(%
\mathbf{r})\mathbf{\alpha }.\mathbf{\tilde{P}+\tilde{P}}^{+}\mathbf{.\alpha }%
V(\mathbf{r})}{2}\right) ^{2}+m^{2}V^{2}(\mathbf{r})}$ $\mathbf{+}\frac{%
\hbar }{4E_{0}}\mathbf{\Gamma }_{0}.\left( mV(\mathbf{r})\mathbf{\Sigma }+%
\frac{\left( \mathbf{\Sigma }.\mathbf{p}\right) \mathbf{p}}{(E_{0}+mV(%
\mathbf{r}))}\right) $ with $\Gamma _{i\gamma }=\varepsilon _{\varrho \beta
\gamma }\Gamma _{i}^{\varrho \beta }(\mathbf{r})$,\ which reads
\begin{equation}
U_{0}\hat{H}_{0}U{}^{+}=\beta \sqrt{P_{i}V^{2}(\mathbf{r})g^{ij}(\mathbf{r}%
)P_{j}+\hbar \varepsilon _{\alpha \beta \gamma }\Gamma _{i}^{\alpha \beta
}h^{i\gamma }\mathbf{\Sigma }.\mathbf{P}+m^{2}V^{2}(\mathbf{r})}\mathbf{+}%
\frac{\hbar }{4E_{0}}\mathbf{\Gamma }_{0}.\left( mV(\mathbf{r})\mathbf{%
\Sigma }+\frac{\left( \mathbf{\Sigma }.\mathbf{p}\right) \mathbf{p}}{%
(E_{0}+mV(\mathbf{r}))}\right)  \label{EO}
\end{equation}

\subsection{Corrections when $\mathbf{P}$ and $\mathbf{R}$ do not commute}

Now, we want to reintroduce the dependence in $\mathbf{R}$, and apply the
method given in \cite{Methode}. Let us however remark that this method has
to be adapted here since our Hamiltonian $E_{0}(\mathbf{P)}$ is not
diagonal, but only block diagonal. Fortunately, given that the block
diagonal part in $E_{0}(\mathbf{P)}$ is only of order $\hbar $, our
formalism still works at the semiclassical level.

As a consequence, to perform our diagonalization procedure, it is sufficient
to apply (at the first order in $\hbar $) the following Foldy-Wouthuysen
transformation
\begin{equation}
U(\mathbf{\tilde{P},R)}=D\left( E+V(\mathbf{R})m+\frac{1}{2}\beta \left( V(%
\mathbf{R})\mathbf{\alpha }.\mathbf{\tilde{P}+\tilde{P}}^{+}\mathbf{.\alpha }%
V(\mathbf{R})\right) +N\right) \mathbf{/}\sqrt{2E\left( E+V(\mathbf{R}%
)m\right) }+X  \label{FW2}
\end{equation}
and then project the transformed Hamiltonian on the positive energy states.
The energy $E$ in Eq.\ref{FW2} is given by expression\ Eq.\ref{EO} where the
parameter $\mathbf{r}$ is replaced by the operator $\mathbf{R}$. All
expressions in $U(\mathbf{\tilde{P})}$ are implicitly assumed to be
symmetrized in $\mathbf{P}$ and $\mathbf{R}$. and the corrective term $X$
must be added to restore the unitarity of $U(\mathbf{\tilde{P})}$ which is
destroyed by the symmetrization. \cite{Methode} shows\textbf{\ }that this
term is expressed as
\begin{equation}
X=\frac{i}{4\hbar }\left[ \mathcal{A}_{P_{l}},\mathcal{A}_{R_{l}}\right] U(%
\mathbf{\tilde{P},R)}
\end{equation}
where we have defined the position and momentum (non projected) Berry phases
$\mathcal{A}_{R}=i\hbar U\nabla _{P}U^{+}$ and $\mathcal{A}_{P}=-i\hbar
U\nabla _{R}U^{+}$. The projection on the positive energy subspace that is
needed to realize the diagonalization leads us to define $\mathbf{A}_{R}=P(%
\mathcal{A}_{R})$ and $\mathbf{A}_{P}=P(\mathcal{A}_{P})$. The resulting
position and momentum operators can thus be written $\mathbf{r}=i\hbar
\partial _{\mathbf{p}}+\mathbf{A}_{R}$ and $\mathbf{p}=\mathbf{P}+\mathbf{A}%
_{P},$ where the explicit computation for the components of the Berry
connections $\mathbf{A}_{R}$ and $\mathbf{A}_{P}$ gives
\begin{equation}
A_{R}^{k}=\hbar c^{2}\frac{\varepsilon ^{\alpha \beta \gamma }h_{\gamma }^{k}%
\tilde{P}_{\alpha }\mathbf{\Sigma }_{\beta }}{2E\left( E+mV(\mathbf{R}%
)\right) }+o\left( \hbar ^{2}\right)   \label{rGB}
\end{equation}

\begin{equation}
A_{P}^{k}=-\hbar c^{2}\frac{\varepsilon ^{\alpha \beta \gamma }\tilde{P}%
_{\alpha }\mathbf{\Sigma }_{\beta }(\mathbf{\nabla }_{R_{k}}\tilde{P}%
_{\gamma })}{2E\left( E+mV(\mathbf{R})\right) }+o\left( \hbar ^{2}\right)
\label{pGB}
\end{equation}
where $E$ is the same as $E_{0}$ above, but now $\mathbf{R}$ is an operator.
Performing our diagonalization process ultimately leads us (after some
computations similar to those presented in \cite{Photon}) to the following
expression for the energy operator
\begin{equation}
\tilde{\varepsilon}\simeq \varepsilon +\hbar \mathbf{B}.\mathbf{\Sigma }%
/2\varepsilon -(\mathbf{A}_{\mathbf{R}}\mathbf{\times p).B/}\varepsilon -%
\frac{1}{2\varepsilon ^{2}}m\hbar \mathbf{\nabla }V(\mathbf{r}).\left(
\mathbf{p}\times \mathbf{\Sigma }\right)
\end{equation}
where
\begin{eqnarray}
\varepsilon  &=&c\sqrt{\left( p_{i}+\hbar \frac{\mathbf{\Gamma }_{i}\left(
\mathbf{r}\right) }{4}.\left( mV(\mathbf{r})\mathbf{\Sigma }+\frac{\left(
\mathbf{\Sigma }.\mathbf{p}\right) \mathbf{p}}{(E+mV(\mathbf{r}))}\right)
\right) V^{2}(\mathbf{r})g^{ij}\left( \mathbf{r}\right) \left( p_{i}+\hbar
\frac{\mathbf{\Gamma }_{i}\left( \mathbf{r}\right) }{4}.\left( mV(\mathbf{r})%
\mathbf{\Sigma }+\frac{\left( \mathbf{\Sigma }.\mathbf{p}\right) \mathbf{p}}{%
(E+mV(\mathbf{r}))}\right) \right) +m^{2}V(\mathbf{r})}  \nonumber \\
&&\mathbf{+}\frac{\hbar }{4E}\mathbf{\Gamma }_{0}.\left( mV(\mathbf{r})%
\mathbf{\Sigma }+\frac{\left( \mathbf{\Sigma }.\mathbf{p}\right) \mathbf{p}}{%
(E+mV(\mathbf{r}))}\right)   \label{Ham}
\end{eqnarray}
with $\Gamma _{i\gamma }\left( \mathbf{r}\right) =\varepsilon _{\alpha \beta
\gamma }\Gamma _{i}^{\alpha \beta }\left( \mathbf{r}\right) $ and the
''magnetotorsion field'' $\mathbf{B}$ is defined through the torsion$%
B_{\gamma }=-\frac{1}{2}P_{\delta }T^{\alpha \beta \delta }\varepsilon
_{\alpha \beta \gamma }$ where $T^{\alpha \beta \delta }=h_{k}^{\delta
}\left( h^{l\alpha }\partial _{l}h^{k\beta }-h^{l\beta }\partial
_{l}h^{k\alpha }\right) +h^{l\alpha }\Gamma _{l}^{\beta \delta }-h^{l\beta
}\Gamma _{l}^{\alpha \delta }$ is the usual torsion for a static metric
(where only space indices in the summations give non zero contributions).
Let us note that in Eq.\ref{Ham} we have neglected the curvature
contributions that appear to be of order $\hbar ^{2}$. Interestingly, this
semi-classical Hamiltonian presents formally the same form as the one of a
Dirac particle in a true external magnetic field \cite{BLIOKH1}\cite{Methode}%
. The term $\mathbf{B}.\mathbf{\sigma }$ is the Stern-Gerlach effect, and
the operator $\mathbf{L}=(A_{R}\mathbf{\times p)}$ is the intrinsic angular
momentum of semiclassical particles. The same contribution $\mathbf{L}$
appears also in the context of the semiclassical behavior of Bloch electrons
(spinless) in an external magnetic field \cite{PIERRE}\cite{NIU} where it
corresponds to a magnetization term. Because of this analogy and since $%
T^{\alpha \beta \delta }$ is directly related to the torsion of space
through $T^{\alpha \beta \delta }=h_{k}^{\delta }h^{i\alpha }h^{j\beta
}T_{ij}^{k}$ we call $\mathbf{B}$ a magnetotorsion field.

Let us also remark that the definition of the Berry curvatures (see below),
allows us to rewrite ultimately the energy $\varepsilon $ in a more compact
form :
\begin{equation}
\varepsilon =c\sqrt{\left( p_{i}-\frac{\mathbf{\Gamma }_{i}\left( \mathbf{r}%
\right) }{2}.\mathbf{\Theta }^{rr}\right) V^{2}(\mathbf{r})g^{ij}\left(
\mathbf{r}\right) \left( p_{i}-\frac{\mathbf{\Gamma }_{i}\left( \mathbf{r}%
\right) }{2}.\mathbf{\Theta }^{rr}\right) +m^{2}V(\mathbf{r})}\mathbf{-}%
\frac{\hbar }{2E}\mathbf{\Gamma }_{0}.\mathbf{\Theta }^{rr}
\end{equation}
where $\Theta ^{rr\gamma }=-\hbar \frac{1}{2E\left( \mathbf{p,r}\right) }%
\left( mV(\mathbf{r})\mathbf{\Sigma }_{\gamma }+\frac{\left( \mathbf{\Sigma }%
^{\delta }\tilde{P}_{\delta }\right) \tilde{P}_{\gamma }}{(E+mV(\mathbf{r}))}%
\right) $ is the ''rescaled'' Berry curvature. This formula clearly shows
that is, the spin connection couples only to the Berry curvature.

The semiclassical Hamiltonian Eq.\ref{Ham} is the main result of this paper.
It contains, in addition to the energy term $\varepsilon $, new
contributions due to the Berry connections. Indeed, the spin couples to the
gravitational field through the magnetotorsion field $\mathbf{B}$ which is
non-zero for a space with torsion. As a consequence, a hypothetical torsion
of space may be revealed through the presence of this coupling.

Let us note at this point that the coupling between torsion and spin has
previously been studied by various authors (\cite{Hehl}, \cite{kerlick},
\cite{oconnel} for example and references therein), but in the context of
general relativity and field equations, whereas the present paper consider
gravitation as a fixed background. Our problematic is thus to describe the
possible effect of torsion on the particle rather than deriving gravity
field equations. From Eqs. \ref{rGB} and \ref{pGB}, we deduce the new (non
canonical) commutations rules :
\begin{eqnarray}
\left[ r_{i},r_{j}\right] &=&i\hbar \Theta _{ij}^{rr} \\
\left[ p_{i},p_{j}\right] &=&i\hbar \Theta _{ij}^{pp} \\
\left[ p_{i},r_{j}\right] &=&-i\hbar g_{ij}+i\hbar \Theta _{ij}^{pr}
\end{eqnarray}
where $\Theta _{ij}^{\alpha \beta }=\partial _{\alpha _{i}}\mathbf{A}_{\beta
_{j}}-\partial _{\beta _{j}}\mathbf{A}_{\alpha _{i}}+[\mathbf{A}_{\alpha
_{i},}\mathbf{A}_{\beta _{j}}]$. An explicit computation shows that
\begin{eqnarray}
\Theta _{ij}^{rr} &=&-\hbar \frac{1}{2\varepsilon ^{3}\left( \mathbf{p,r}%
\right) }\left( mV(\mathbf{r})\mathbf{\Sigma }_{\gamma }+\frac{\left(
\mathbf{\Sigma }^{\delta }\tilde{P}_{\delta }\right) \tilde{P}_{\gamma }}{%
(E+mV(\mathbf{r}))}\right) \varepsilon ^{\alpha \beta \gamma }h_{\alpha
}^{i}h_{\beta }^{j} \\
\Theta _{ij}^{pp} &=&-\hbar \frac{1}{2\varepsilon ^{3}\left( \mathbf{p,r}%
\right) }\left( mV(\mathbf{r})\mathbf{\Sigma }_{\gamma }+\frac{\left(
\mathbf{\Sigma }^{\delta }\tilde{P}_{\delta }\right) \tilde{P}_{\gamma }}{%
(E+mV(\mathbf{r}))}\right) \nabla _{r_{i}}\tilde{P}_{\alpha }\nabla _{r_{j}}%
\tilde{P}_{\beta }\varepsilon ^{\alpha \beta \gamma } \\
&&+\frac{\hbar }{2\varepsilon ^{3}\left( \mathbf{p,r}\right) }m\left[ \nabla
_{r_{i}}V(\mathbf{r})\left( \left( \mathbf{\Sigma }\times \mathbf{\tilde{P}}%
\right) .\nabla _{r_{j}}\mathbf{\tilde{P}}\right) -\nabla _{r_{j}}V(\mathbf{r%
})\left( \left( \mathbf{\Sigma }\times \mathbf{\tilde{P}}\right) .\nabla
_{r_{i}}\mathbf{\tilde{P}}\right) \right] \\
\Theta _{ij}^{pr} &=&\hbar \frac{1}{2\varepsilon ^{3}\left( \mathbf{p,r}%
\right) }\left( mV(\mathbf{r})\mathbf{\Sigma }_{\gamma }+\frac{\left(
\mathbf{\Sigma }^{\delta }\tilde{P}_{\delta }\right) \tilde{P}_{\gamma }}{%
(E+mV(\mathbf{r}))}\right) \nabla _{r_{i}}\tilde{P}_{\alpha }h_{\beta
}^{j}\varepsilon ^{\alpha \beta \gamma } \\
&&-\frac{\hbar }{2\varepsilon ^{3}\left( \mathbf{p,r}\right) }m\nabla
_{r_{i}}V(\mathbf{r})\left( \mathbf{\Sigma }\times \mathbf{\tilde{P}}\right)
_{j}
\end{eqnarray}
where the couple $\left( \mathbf{P,R}\right) $ has been replaced by $\left(
\mathbf{p,r}\right) $ in $\mathbf{\tilde{P}}$. We will need also
\begin{eqnarray}
\Theta _{ij}^{r\Sigma }=\left[ r_{i},\Sigma _{j}\right] &=&i\hbar c^{2}\frac{%
-p_{j}\mathbf{\Sigma }_{i}+\mathbf{p.\Sigma }\delta _{ij}}{\varepsilon
\left( \mathbf{p,r}\right) \left( \varepsilon \left( \mathbf{p,r}\right) +mV(%
\mathbf{r})\right) }+o\left( \hbar ^{2}\right) \\
\Theta _{ij}^{p\Sigma }=\left[ p_{i},\Sigma _{j}\right] &=&-i\hbar c^{2}%
\frac{-p_{j}\mathbf{\Sigma }_{l}+\mathbf{p.\Sigma }\delta _{lj}}{\varepsilon
\left( \mathbf{p,r}\right) \left( \varepsilon \left( \mathbf{p,r}\right) +mV(%
\mathbf{r})\right) }h_{l}^{\gamma }\mathbf{\nabla }_{r_{i}}\tilde{p}_{\gamma
}+o\left( \hbar ^{2}\right)
\end{eqnarray}
Ultimately, using the commutation relations at hand plus the Hamiltonian, we
can easily derive the equations of motion for the electron in a static
gravitational field. As explained in \cite{Methode}, the chosen dynamical
variables are $\mathbf{r}$ and $\mathbf{p}$. This choice comes quite
naturally when considering the projection in the diagonalization process. It
is also this choice that allows to take into account for the spin Hall
effect for the photon, for example. We thus obtain
\begin{eqnarray}
\mathbf{\dot{r}} &=&\left( 1-\Theta ^{pr}\right) \nabla _{\mathbf{p}}\tilde{%
\varepsilon}+\mathbf{\dot{p}\times }\Theta ^{rr}+\frac{i}{\hbar }\nabla _{%
\mathbf{\Sigma }}\tilde{\varepsilon}.\Theta _{ij}^{r\Sigma }  \nonumber \\
\mathbf{\dot{p}} &=&-\left( 1-\Theta ^{pr}\right) \nabla _{\mathbf{r}}\tilde{%
\varepsilon}+\mathbf{\dot{r}}\times \Theta ^{pp}+\frac{i}{\hbar }\nabla _{%
\mathbf{\Sigma }}\tilde{\varepsilon}\Theta _{ij}^{p\Sigma }  \label{EQR}
\end{eqnarray}
However, these equations are incomplete per se since they involve the
initial spin matrix $\hbar \mathbf{\Sigma }$ that is not conserved through
the dynamical evolution. To compute fully the dynamics for the electron at
the first order in $\hbar $, these equations have consequently to be
completed with the dynamics of the spin matrix $\hbar \mathbf{\Sigma }$ at
the lowest order
\begin{eqnarray*}
\hbar \dot{\Sigma} &=&\frac{i}{\hbar }\left[ \tilde{\varepsilon},\hbar
\Sigma \right] =p^{i}\left( \frac{m}{4\varepsilon }V(\mathbf{r})\left(
\mathbf{\Gamma }_{i}\left( \mathbf{r}\right) \times \mathbf{\Sigma }\right) +%
\frac{\hbar }{4}\frac{\left( \mathbf{\Gamma }_{i}\left( \mathbf{r}\right) .%
\mathbf{p}\right) \left( \mathbf{p}\times \mathbf{\Sigma }\right) }{%
\varepsilon +mV(\mathbf{r})}\right) -\frac{\hbar }{4}\left( \mathbf{\Gamma }%
_{0}\mathbf{+}\frac{\left( \mathbf{\Gamma }_{0}\times \mathbf{P}\right)
\times \mathbf{P}}{E_{0}^{2}}\right) \times \mathbf{\Sigma } \\
&&-\frac{m\hbar }{2\varepsilon ^{2}}\left( \mathbf{B}\times \mathbf{\Sigma }%
\right) -\hbar \frac{\mathbf{p\mathbf{B}}\left( \mathbf{p\times \Sigma }%
\right) }{2\varepsilon ^{2}\left( \varepsilon +mV(\mathbf{R})\right) }-\frac{%
m\hbar }{\varepsilon ^{2}}\left[ \left( \mathbf{\nabla }V(\mathbf{r})\times
\mathbf{p}\right) \mathbf{\times \Sigma }\right]
\end{eqnarray*}

The velocity equation contains in particular an anomalous velocity term $%
\mathbf{\dot{p}\times }\Theta ^{rr}$, which has been described in several
other circonstances. It is for instance responsible for the intrinsic spin
Hall effect of Dirac electrons in electromagnetic fields \cite{ALAIN,BLIOKH1}%
. In semiconductor, SO coupling being greatly enhanced with respect to the
vacuum case, this anomalous velocity drastically modifies the transport
properties of the charges \cite{ZHANG}. A similar anomalous velocity appears
also in the context of spinless electrons in magnetic Bloch bands \cite
{PIERRE}\cite{NIU}. Here for the first time we predict an anomalous velocity
contribution for an electron propagating in a static arbitrary graviational
background. Note that this effect is present independently of the existence
of a torsion of space.

The force equation presents the dual expression $\mathbf{\dot{r}}\times
\Theta _{pp}$ of the anomalous velocity which is a kind of Lorentz force
which being of order $\hbar $ does not influence the velocity equation at
order $\hbar $. It is interesting to remark that similar equations of motion
with dual contributions $\mathbf{\dot{p}\times }\Theta _{rr}$ and $\mathbf{%
\dot{r}}\times \Theta _{pp}$ were predicted for the wave-packets dynamics of
spinless electrons in crystals subject to small perturbations \cite{NIU}.

Within our approach we can easily treat the ultrarelativistic limit $%
mc^{2}\rightarrow 0$ which by simplifying all expressions gives a better
feeling of the physics.

\subsection{Ultrarelatvistic limit}

In the ultrarelativistic limit $m\rightarrow 0$, one recovers the same kind
of Hamiltonian derived in \cite{Photon} for the photon case. Indeed one
readily obtain
\begin{equation}
\widetilde{\varepsilon }\simeq \varepsilon +\frac{\lambda }{4}\frac{\mathbf{%
p.\Gamma }_{0}}{p}+\frac{\lambda g_{00}}{2\varepsilon }\frac{\mathbf{B}.%
\mathbf{p}}{p}  \label{NUMBER}
\end{equation}
where $\varepsilon =c\sqrt{\left( p_{i}+\frac{\lambda }{4}\frac{\Gamma _{i}(%
\mathbf{r}).\mathbf{p}}{p}\right) g^{ij}g_{00}\left( p_{j}+\frac{\lambda }{4}%
\frac{\Gamma _{j}(\mathbf{r}).\mathbf{p}}{p}\right) }$ with $\lambda =%
\mathbf{p.\Sigma /}p$. Interestingly this energy can be expressed in terms
of the helicity and not in term of $\Sigma $. In fact the terms $\frac{i}{%
\hbar }\nabla _{\mathbf{\Sigma }}\tilde{\varepsilon}.\Theta _{ij}^{r\Sigma }$
and $\frac{i}{\hbar }\nabla _{\mathbf{\Sigma }}\tilde{\varepsilon}\Theta
_{ij}^{p\Sigma }$ in Eq.\ref{EQR} recombines with the gradient energy
contributions which allows us to rewrite the equation of motion under the
following form:

\begin{eqnarray}
\mathbf{\dot{r}} &=&\left( 1-\Theta _{pr}\right) \nabla _{\mathbf{p}}\tilde{%
\varepsilon}+\mathbf{\dot{p}\times }\Theta _{rr}  \nonumber \\
\mathbf{\dot{p}} &=&-\left( 1-\Theta _{pr}\right) \nabla _{\mathbf{r}}\tilde{%
\varepsilon}+\mathbf{\dot{r}}\times \Theta _{pp}  \label{Eqmotion}
\end{eqnarray}
where in $\nabla _{\mathbf{p}}\tilde{\varepsilon}$ we must consider $\mathbf{%
p}$ and $\mathbf{\Sigma }$ as independent variables. The term $\mathbf{\dot{p%
}\times }\Theta _{rr}$ causes an additional displacement of electrons of
distinct helicity in opposite directions orthogonally to the ray. In
comparison to the usual velocity $\mathbf{\dot{r}}=\nabla _{\mathbf{p}}%
\tilde{\varepsilon}$ $\sim c$, the anomalous velocity term $\mathbf{v}%
_{\perp }$ is obviously small, its order $v_{\perp }^{i}\sim c\widetilde{%
\lambda }\nabla _{r^{j}}g^{ij}$ being proportional to the wave length $%
\widetilde{\lambda }$.

To complete the dynamical description notice that at the leading order the
helicity $\lambda $ is not changed by the unitary transformation which
diagonalizes the Hamiltonian so that it can be written $\lambda =\hbar
\mathbf{p}.\mathbf{\Sigma }/p$. After a short computation one can check that
the helicity is always conserved
\begin{equation}
\frac{d}{dt}\left( \frac{\hbar \mathbf{p.\Sigma }}{p}\right) =0
\end{equation}
for an arbitrary static gravitational field independently of the existence
of a torsion of space.

\section{The symmetric gravitational field}

A typical example of such a metric is the Schwarzschild space-time in
isotropic coordinates. This case, studied in a different manner in \cite
{OBUKHOV}and \cite{SILENKO}. The Hamiltonian is given by, received a full
independent treatment within our formalism in \cite{Methode}. We thus
present directly the needed results completed with the spin matrix dynamics.
For a symmetric metric one has $\mathbf{B.p=\Gamma }_{0}=0$ and the
semiclassical Hamiltonian can take the following form \cite{SILENKO}

\begin{equation}
H_{0}=\frac{1}{2}\left( \alpha .PF(\mathbf{R})+F(\mathbf{R})\alpha .P\right)
+\beta mV(\mathbf{R})
\end{equation}
corresponding to the metric $g_{ij}=\delta _{ij}\left( \frac{V(\mathbf{R})}{%
F(\mathbf{R})}\right) ^{2}$, $g_{i0}=0$ and $g_{00}=V^{2}(\mathbf{R})$.\
Similar computations to the ones performed in the previous section lead to
the following expressions for the dynamical variables and the diagonalized
Hamiltonian
\begin{eqnarray}
\mathbf{r} &=&\mathbf{R+}\mathcal{A}_{R}^{+}=\mathbf{R-}\hbar \frac{F^{2}(%
\mathbf{R})\mathbf{\Sigma }\times \mathbf{P}}{2E(E+mV(\mathbf{R}))} \\
\mathbf{p} &=&\mathbf{P+}\mathcal{A}_{P}^{+}=\mathbf{P}
\end{eqnarray}
\begin{equation}
\varepsilon \left( \mathbf{p,r}\right) =\sqrt{F^{2}(\mathbf{r})\mathbf{P}%
^{2}+\mathbf{P}^{2}F^{2}(\mathbf{r})+mV^{2}(\mathbf{r})}-\frac{F^{3}(\mathbf{%
\ r})}{2E^{2}}m\hbar \mathbf{\nabla }\phi (\mathbf{r}).\left( \mathbf{P}%
\times \mathbf{\Sigma }\right)
\end{equation}
with $\phi =\frac{V}{F}$. The commutators of the dynamical variables define
the Berry curvatures $\left[ r_{i},r_{j}\right] =i\hbar \Theta _{ij}^{rr},$ $%
\left[ P_{i},r_{j}\right] =-i\hbar g_{ij}+i\hbar \Theta _{ij}^{pr}$ and $%
\left[ P_{i},P_{j}\right] =0,$ with
\begin{eqnarray}
\Theta _{ij}^{rr} &=&-\frac{\hbar F^{3}(\mathbf{r})\varepsilon ^{ijk}}{%
2\varepsilon ^{3}\left( \mathbf{P,r}\right) }\left( m\phi (\mathbf{r})%
\mathbf{\Sigma }_{k}+\frac{F(\mathbf{r})\left( \mathbf{\Sigma .P}\right)
\mathbf{P}_{k}}{\varepsilon \left( \mathbf{P,r}\right) +mV(\mathbf{r})}%
\right) \\
\Theta _{ij}^{pr} &=&-\frac{\hbar F^{3}(\mathbf{r})}{2\varepsilon ^{3}\left(
\mathbf{P,r}\right) }m\nabla _{i}\phi (\mathbf{r})\left( \mathbf{\Sigma }%
\times \mathbf{P}\right) _{j} \\
\Theta _{ij}^{pp} &=&0
\end{eqnarray}
and $\Theta _{ij}^{r\Sigma }$ being unchanged and $\Theta _{ij}^{p\Sigma }=0$%
. One can check, after developing $\mathbf{r}$ as a function of $\mathbf{R}$
and the Berry phase, that our Hamiltonian coincides with the one given in
\cite{SILENKO} when considering the semiclassical limit (order $\hbar )$.
This also confirms the validity of the Foldy Wouthuysen approach asserted in
\cite{SILENKO}. The absence of dipole spin gravity interaction, in
opposition with the transformation proposed in \cite{OBUKHOV}, results only
from a different choice of diagonalization matrix.

Let us ultimately remark that, despite some similarities, our approach is
different from the one developed in \cite{SILENKO}. Actually, this last
paper considers an approximated Foldy Wouthuysen transformation at the first
order in a (weak) gravitational field. On the contrary, our approach, even
if semiclassical, allows to consider a Foldy Wouthuysen transformation for
an arbitrary gravitational field.

To conclude this paragraph, we can easily derive the equations of motion
from the commutation relations
\begin{eqnarray}
\mathbf{\dot{r}} &=&\nabla _{\mathbf{P}}\tilde{\varepsilon}-\mathbf{\dot{P}%
\times }\Theta ^{rr}+\frac{i}{\hbar }\nabla _{\mathbf{\Sigma }}\tilde{%
\varepsilon}.\Theta _{ij}^{r\Sigma } \\
\mathbf{\dot{P}} &=&-\nabla _{\mathbf{r}}\tilde{\varepsilon}+\nabla _{%
\mathbf{r}}\varepsilon .\Theta ^{pr} \\
&=&-\nabla _{\mathbf{r}}\tilde{\varepsilon}+\nabla _{\mathbf{r}%
_{j}}\varepsilon \frac{\hbar F^{3}(\mathbf{r})}{2\varepsilon ^{3}\left(
\mathbf{P,r}\right) }m\nabla _{i}\phi (\mathbf{r})\left( \mathbf{P}\times
\mathbf{\Sigma }\right) _{j}
\end{eqnarray}
\[
\hbar \mathbf{\dot{\Sigma}=}\frac{mV(\mathbf{r})\hbar }{\varepsilon
(\varepsilon +mV(\mathbf{r}))}\mathbf{\Sigma \times }\left( \mathbf{\nabla }%
V(\mathbf{r})\mathbf{\times P}\right) -\frac{\hbar }{\varepsilon }\mathbf{%
\Sigma \times }\left( \mathbf{\nabla }F(\mathbf{r})\mathbf{\times P}\right)
\]
Although the first equation differs from the one ontained in \cite{SILENKO}
due to a different choice of dynamical position variable, the last two
equations reduce to \cite{SILENKO}, in the case of weak fields, :
\begin{eqnarray*}
\mathbf{\dot{P}} &=&-\frac{m^{2}}{\varepsilon }\mathbf{\nabla }V(\mathbf{r})-%
\frac{p^{2}}{\varepsilon }\mathbf{\nabla }F(\mathbf{r})+\frac{m\hbar }{%
2\varepsilon (\varepsilon +m)}\mathbf{\nabla }\left( \mathbf{\Sigma .}\left(
\mathbf{\nabla }V(\mathbf{r})\mathbf{\times P}\right) \right) -\frac{\hbar }{%
2\varepsilon }\mathbf{\nabla }\left( \mathbf{\Sigma .}\left( \mathbf{\nabla }%
F(\mathbf{r})\mathbf{\times P}\right) \right) \\
\hbar \mathbf{\dot{\Sigma}} &=&\frac{m\hbar }{\varepsilon (\varepsilon +m)}%
\mathbf{\Sigma \times }\left( \mathbf{\nabla }V(\mathbf{r})\mathbf{\times P}%
\right) -\frac{\hbar }{\varepsilon }\mathbf{\Sigma \times }\left( \mathbf{%
\nabla }F(\mathbf{r})\mathbf{\times P}\right)
\end{eqnarray*}
where here $\varepsilon =\sqrt{p^{2}+m^{2}}.$

Let us stress again that our dynamical position variables are not the
canonical variables $\mathbf{R}$ chosen in \cite{SILENKO}, but the non
commutative ones $\mathbf{r}$. As a consequence, our equations obviously
differ from theirs by the role of the Berry curvature, even if the
Hamiltonians are formally identical.

\section{Conclusion}

The semiclassical limit for Dirac particles interacting with a static
gravitational field was investigated. We found a Foldy-Wouthuysen
transformation which diagonalizes at the semiclassical order the Dirac
equation for an arbitrary static space-time metric (not necessarily
symmetrical contrary to previous works). The main results are the following.

First, the energy includes, through the Berry phases, a coupling between the
torsion of the space and the spin of the particle. Second, we have been able
to derive the semi classical equations of motion for the electron. Our set
of dynamical variables is non commutative and this choice induces new
corrections to the dynamical equations that could be relevant in a strong
gravitational field. Ultimately, we retrieve the particular case of a
symmetric static spacetime metric \cite{SILENKO}, and confirms the
generality of our approach

\end{document}